\documentstyle[prl,aps,epsf,graphicx,preprint]{revtex}

\nonumber

\begin{document}

\vskip 1cm

\title{Off-equilibrium dynamics in the energy landscape of a simple model glass}
\author{ 
        R.~Di Leonardo$^{1}$,
        L.~Angelani$^{2}$,
        G.~Parisi$^{3}$,
        G.~Ruocco$^{3}$, 
	A.~Scala$^{3}$, and 
	F.~Sciortino$^{3}$ 
        }
\maketitle
\centerline{ 
         $^1$
         Universit\`a di L'Aquila
         and INFM,
         I-67100, L'Aquila, Italy.
		}

\centerline{ 
         $^2$
         Universit\`a di Trento and INFM, I-38050, Povo, Trento, Italy.\\
		}
\centerline{ 
         $^3$
         Universit\`a di Roma {\it La Sapienza} and INFM, I-00185,
          Roma, Italy.
        	}
\date{}

\medskip

\begin{center}

\begin{abstract}
The aging dynamics of a simple model glass is numerically investigated
observing how it takes place in the potential energy landscape $V$.
Partitioning the landscape in basins of minima of $|\nabla V|^2$, 
we are able to elucidate some interesting topological 
properties of the aging process.
The main result is the characterization of the long time behavior as a jump dynamics 
between basins of attraction of minima.
Moreover we extract some information about the landscape itself, determining 
quantitatively few parameters describing it, 
such as the mean energy barrier value and the 
mean square distance between adjacent minima.

\end{abstract}
\end{center}

\vskip 0.7cm

The specific way a glass or a supercooled liquid explores its potential energy 
landscape at different temperatures and in different dynamical regimes is of great  
importance to understand many properties of these systems 
(Stillinger, 1995;
Debenedetti and Stillinger, 2001). 
Only recently the investigation of the multidimensional Potential Energy Surface (PES)
as a function of particle coordinates has become computational 
achievable, and many numerical works have been devoted to study the PES of 
model liquids and glasses. 
The relationship between the topological properties of the 
landscape and the way the system moves in it, constitutes a very promising viewpoint 
to better understand the thermodynamics and dynamics of supercooled liquids and 
glassy systems (Kurchan and Laloux, 1996).

Only few studies have investigated the aging dynamics from the energy 
landscape point of view (Angelani {\it et al.} 2001;
Kob {\it et al.} 2000; Mossa {\it et al.} 2001).
The study of the off-equilibrium dynamics in glassy systems, 
and in particular in spin glasses, has allowed 
to obtain important information on the phase-space properties 
of the systems itself 
and to formulate off-equilibrium relations generalizing the usual equilibrium ones 
- generalized fluctuation-dissipation relation 
(Cugliandolo, and Kurchan 1993;
Bouchaud {\it et al.} 1997) - 
with the introduction of an effective temperature
(Cugliandolo {\it et al.} 1997).
The investigation of the off-equilibrium properties seems 
to indicate the correctness of the conjecture stating 
the similarity of structural glasses with some spin glass model 
- spin glasses with one step replica symmetry breaking
(Kirkpatrick and Thirumalai, 1987;
Kirkpatrick and Wolynes, 1987;
Franz and Parisi, 1997).
Recently the numerical investigation of aging in structural glasses has 
allowed us to bridge these two different disordered systems
(Parisi  1997; Barrat and Kob 1999; 
Di Leonardo {\it et al.} 2000;
Sciortino and Tartaglia 2001;).
Here we report a detailed analysis of the aging in the energy landscape 
for a simple model glass, a monatomic Lennard-Jones system modified in such a 
way as to avoid crystallization.

The way the off-equilibrium dynamics takes place in the potential energy landscape $V$ 
(viewed as a function of the $3N$ coordinates of the $N$ particles)
is investigated by mapping the true dynamics in the configuration space on to 
the ``slow'' points of $V$, i.e. the points in which $|\nabla V|^2$ is minimum 
and the forces are reduced by several orders of magnitude
(we call them {\it saddles} hereafter, even if, besids true saddles, they include also 
inflection points - Doye and Wales 2001; Angelani {\it et al.}, in preparation).
The basins of attraction of these points constitute a partition of the configuration space.
This partition of the landscape is an extension of the partition in basins of minima
proposed by Stillinger and Weber (1982, 1984) many years ago, 
including in our case not only the minima but also the {\it saddles} of any order.
The new partition has been very useful to elucidate many interesting features of supercooled 
simple liquids, the main result being the identification of the mode-coupling temperature $T_{MCT}$
(G\"otze, 1999)
as the temperature at which the system has a crossover between a dynamics between 
basins of {\it saddles} for $T > T_{MCT}$ to a dynamics between basins of minima for $T < T_{MCT}$
(Angelani {\it et al.} 2000; Broderix {\it et al.} 2000; Scala {\it et al.} 2001).
We use this new partition in the analysis of the off-equilibrium dynamics 
of a monatomic Lennard-Jones system with $N=256$ particles. 
The particles interact via the usual $6-12$ Lennard-Jones potential modified
in such a way as to avoid the crystalization occuring in standard LJ:
$V=V_{LJ} + \delta V$, where $\delta V$ is a small many-body term that inhibits cristalization
(it depends on the static structure factor, see 
Di Leonardo {\it et al.} (2000) for details).
The relevant quantitites are determined during the time evolution of the system
after a sudden crunch (density jump at fixed temperature) from liquid to glassy phase 
(from $\rho_0=0.95$ to $\rho_1=1.24$ at fixed $T_0=0.5$, where the glass transition 
temperature is $T_g \sim 1.4$ at the final density $\rho_1$ 
- all the quantities are expressed in standard Lennard-Jones unit).
The evolution of the system is generated up to time $t=3\cdot 10^4$ by means of standard
isothermal MD with time step $\tau=0.01$.
The measured quantities have been averaged over $80$ samples.
For each MD trajectory we have quenched a set of logarithmically equally
spaced configurations to the corresponding minima and {\it saddles} 
following the steepest descent path respectively on the potential
energy surface $V$ in the former case 
and on the auxiliary potential $W=\frac{1}{2}|\nabla V|^2$
(Weber and Stillinger 1985) in the latter.
In this way, during the time evolution, we are able to follow the behavior of the sampled 
relevant points of the landscape: {\it saddles} and minima (besides the usual instantaneous configurations).

In order to elucidate how the system explores the landscape during the off-equilibrium dynamics,
we have first focused on the energy values.
We have measured the potential energies as a function of time 
(times are measured from crunch) along the dynamical trajectory:
the energy of the instantaneous configurations, 
the energy of the {\it saddles} and the energy of the minima.
The data are reported in Fig. 1 in a logarithmic scale.
It is evident that the instantaneous and saddle energies have a rather sharp transition 
at time $t_c \sim 20$. 
The meaning of this crossover will be clear soon, analyzing more carefully another 
topological quantity related to the landscape, the order of sampled {\it saddles}.
We only notice here that for times less than $t_c$ the instantaneous energy follows the 
energy of {\it saddles}, and the energy of minima remains nearly constant in this region
(we note this flat time-region corresponds to the constant energy region observed at high temperature
in the analysis of minima in equilibrated liquids),
while for times below $t_c$ all the energies (instantaneous, {\it saddles} and minima) have 
a similar almost logarithmic behavior, 
indicating that here the time evolution is driven by the inherent dynamics between minima: 
$e(t) \propto \gamma \log_{10}(t)$,  with $\gamma \sim - 0.025$.

Another remarkable point that can be observed in Fig. 1 is the similarity between 
the energy of the {\it saddles} and the difference between the instantaneous energy and the constant 
term $\frac{3}{2}T_0$.
This seems to indicate that the non-diffusive degrees of freedom give an harmonic contribution 
over the energy of {\it saddles}, slowly approaching their equilibrium value.

A better understanding of the processes that are taking place is obtained analyzing the number of negative
directions in the landscape near the sampled {\it saddles}. 
We define the saddle order $n_s$ as the number of negative curvatures at the saddle point, 
i.e. the fraction of negative eigenvalues of the Hessian of $V$. 
The time evolution of $n_s$ during aging (Fig. 2) allows us to clarify the meaning 
of the crossover time $t_c$:
it marks the time at which the system begins to visit extensively the minima, 
and the mean order $n_s$ falls below the value $n_s = 1$.

In the analysis of equilibrium properties of supercooled liquids,
the quantity $n_s$ vs. temperature allowed us to give a topological interpretation
of the mode-coupling temperature as the temperature at which a transition takes place,
from a dynamics between {\it saddles} to a dynamics between minima 
(Angelani {\it et al.} 2000).
In a similar way, in the present study of the out-of-equilibrium properties of a simple glass,
the same quantity $n_s$, now plotted vs. time $t$, allows us to interpret the 
time $t_c$ as marking the transition between the same two regimes: at short times, below $t_c$,
the system moves inside basins of attraction of {\it saddles} with mean order $n_s(t) > 1$ decreasing in time,
while at longer times, $t>t_c$, the system starts to visit basins of minima 
traveling among them through low order {\it saddles}. 
The fact that the mean order $n_s$ doesn't vanish for times $t>t_c$, explains the observed energy difference
between the energy of {\it saddles} and that of minima.

Analyzing how the system explore its potential energy landscape during the aging dynamics
we can try to infer some topological characteristic of the landscape.
Matching the information about the order and energy of {\it saddles} and the energy of minima, we are able to 
evidence the possible existence of a typical energy barrier in the landscape.
As reported in a previous work investigating the equilibrium supercooled dynamics
(Angelani {\it et al.} 2000),
also in the aging regime we discover a very simple topology of the landscape:
{\it saddles} of order $n_s$ lie above the underlying minimum 
by an energy elevation proportional to $n_s$:
\begin{equation}
\label{en_or}
e_s (n_s) - e_m = \frac{\Delta E}{N} \cdot n_s \ .
\end{equation}
In Fig. 3 the left hand side of Eq. $1$ is reported as a function of saddle order $n_s$.
The open circles are the data obtained from previous figures, in which the minima are defined 
by a steepest descent path from the instantaneous configuration.
However, using a different definition of underlying minima below a saddle (by quenching the saddle
instead of instantaneous configuration - see the text below) 
one observes a very similar behaviour (full diamonds in Fig. 3).
This finding supports the picture of a very simple landscape with only a single energy barrier
parameter $\Delta E$ which represents the energy gap between a saddle 
of order $n_s$ and a saddle of order $n_s+1$:
\begin{equation}
\label{barr}
\Delta E \sim 10 \ .
\end{equation}
Similar results were found in the supercooled equilibrium case, where a value of $\Delta E \sim 3.6$ 
was obtained for a less dense system $\rho = 1$ (in our case $\rho = 1.24$).

Another interesting topological information about the landscape is the valuation of the mean 
distance $d^2$ between adjacent minima. In order to obtain this information we have determined, for each 
saddle point sampled during the dynamics, two underlying minima
(here the term {\it underlying} means minima reached by a steepest descent path from a saddle configuration, 
weakly perturbed in order to escape from the instable point), 
$\{ {\bf r}_i^{(1)} \}$ and  $\{{\bf r}_i^{(2)}\}$, reached by quenching the saddle.
In this case the above definition of minima, as those reached by {\it saddles}, seems to be the more appropiate, 
due to the fact that the minima sampled during the dynamics starting from instantaneous configurations
could be not directly related to the sampled {\it saddles}. 
Instead a direct quench from {\it saddles} assures the close relation in configuration space between the 
two points (we note that in the valuation of energies the two method lead to the same results, 
indicating that the energies are less sensitive to landscape details).
The quantity $d^2$ defined as
\begin{equation}
d^2 =\frac{1}{N}\sum_i\langle 
|{\mathbf r}_i^{(1)}-{\mathbf r}_i^{(2)}|^2\rangle \ ,
\end{equation} 
is reported in Fig. 4 as a function of the saddle order $n_s$.
It is evident a linear relation (line in Fig. 4) between the two quantities, 
suggesting that the descent path from a saddle of order $n_s$ towards the 
underlying minima can be interpreted as a random walk among {\it saddles}:
at each step the order of the saddle decreases by 1 and the mean square distance 
increases by a constant value $d^2_0$
(this characteristic, together with the single barrier energy parameter $\Delta E$, 
is incorporated in the simple trigonometric model as defined by Madam and Keyes (1993)).
The value obtained for the typical square distance between minima is:
\begin{equation}
d^2_0 \sim 0.02 \ .
\end{equation}
We note that this finding and the previous one concerning the mean barrier value (see Eq. $2$), 
although obtained in the specific aging regime, are supposed to be properties of the landscape
itself, independent on the different way the system explores it.
Indeed the same characteristics were obtained by Angelani {\it et al.} (2000) who analysed 
the equilibrium dynamics in the supercooled
regime (a quantitative match between 
the obtained values is not feasible, as the densities of the systems are different in the two cases).

In conclusion, the topological investigation of the energy landscape of a simple model glass during the 
aging dynamics has allowed us to clarify many interesting properties, shortly resumed here after:
\begin{itemize}
\item after a fast short time behavior, characterized by a dynamics between {\it saddles}, the system starts to explore
minima decreasing logarithmically its energy (this latter is the interesting slow aging dynamics);
\item the instantaneous energies lie above the saddle energies by the harmonic term $\frac{3}{2} T$, 
indicating that the fast harmonic degrees of freedom equilibrate over the 
slow diffusive coordinates (related to the negative directions of the sampled {\it saddles});
\item the topology of the landscape seems to be highly regular, with only a single energy barrier parameter 
$\Delta E \sim 10$ and a single square distance between adjacent minima $d_0^2 \sim 0.02$.
\end{itemize}

\newpage

\centerline{\large REFERENCES}

\vskip 0.7cm

\noindent  Angelani L., Di~Leonardo R., Ruocco G., Scala A. , and Sciortino F.,
2000, Phys. Rev. Lett. {\bf 85}, 5356.

\noindent  Angelani L., Di~Leonardo R., Ruocco G., Scala A. , and Sciortino F.,
in preparation.

\noindent  Angelani L., Di~Leonardo R., Ruocco G., and Parisi G.,
2001, Phys. Rev. Lett. {\bf 87}, 055502.

\noindent Barrat J.L., and W. Kob, 1999, Europhys. Lett. {\bf 46}, 637.

\noindent Bouchaud J.P., Cugliandolo L.F., Kurchan J., and Mezard M., 1997,
       {\it Out of equilibrium dynamics in spin glasses and the
       other glassy systems, } in
       {\it Spin Glasses and Random Fields, } (World Scientific, Singapore).

\noindent Broderix K., Batthacharya K.K., Cavagna A., Zippelius A., and Giardina I.,
2000, Phys. Rev. Lett. {\bf 85}, 5360.

\noindent Cugliandolo L.F., and Kurchan J., 1993, Phys. Rev.
Lett. {\bf 71}, 173.

\noindent Cugliandolo L.F., Kurchan J., and Peliti K. 1997, 
Phys. Rev. E {\bf 55}, 3898.

\noindent Debenedetti P.G., and Stillinger F.H., 2001, Nature, {\bf 410}, 259.

\noindent Di~Leonardo R., Angelani L., Parisi G., and  Ruocco G., 2000
Phys. Rev. Lett. {\bf 84}, 6054.

\noindent Doye J.P.K and Wales D.J., cond-mat/0108310.

\noindent Franz S., and Parisi G., 1997, Phys. Rev. Lett. {\bf 79}, 2486.

\noindent G\"otze W., 1999,  { J. Phys.: Condens. Matter} {\bf 11}, A1.

\noindent Kirkpatrick T.R., and Thirumalai D., 1987, Phys. Rev.
B {\bf 36}, 5388.

\noindent Kirkpatrick T.R., and Wolynes P.G., 1987, Phys. Rev.
B {\bf 35}, 3072.

\noindent Kob W., Sciortino F., and Tartaglia P., 2000, Europhys. Lett. {\bf 49}, 590.

\noindent Kurchan J., and Laloux L., 1996, J. Phys. A {\bf 29}, 1929.
      
\noindent Madam B., and Keyes T., 1993, J. Chem. Phys. {\bf 98}, 3342.

\noindent Mossa S. et al., 2001, Phil. Mag. B. {\bf 82}, in press.

\noindent Parisi G., 1997, Phys. Rev. Lett. {\bf 79}, 3660.

\noindent Scala A. et al., 2001, Phil. Mag. B. {\bf 82}, in press.

\noindent Sciortino F.,and Tartaglia P., 2001, Phys. Rev. Lett. {\bf 86}, 107.

\noindent Stillinger F.H., 1995, Science {\bf 267}, 1935.

\noindent  Stillinger F.H., and Weber T.A., 1982, Phys. Rev. A {\bf 25}, 978.

\noindent Stillinger F.H., and Weber T.A., 1984,  Science {\bf 225}, 983.

\noindent Weber T.A., and Stillinger F.H., 1985, Phys. Rev. B {\bf 31}, 1954.

\newpage

\begin{figure}[hbt]
\centering
\includegraphics[width=.5\textwidth,angle=-90]{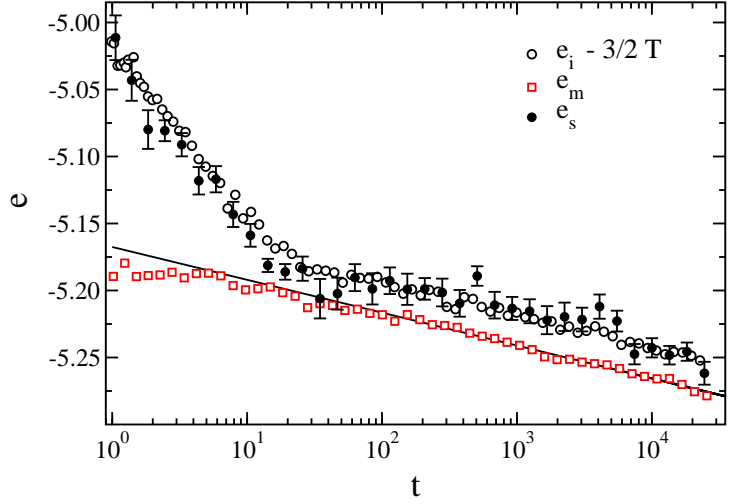}
\caption{
Energies (per particle) during the aging dynamics. 
Instantaneous potential energy $e_i$ subtracted by the constant term $3/2 T$ (open circles),
energy of minima $e_m$ (open squares) and energy of {\it saddles} (full circles).
The solid line fits the long time data, with slope $\gamma \sim - 0.025$.}
\label{fig1}
\end{figure}

\vspace{-1.cm}

\begin{figure}[hbt]
\centering
\includegraphics[width=.5\textwidth,angle=-90]{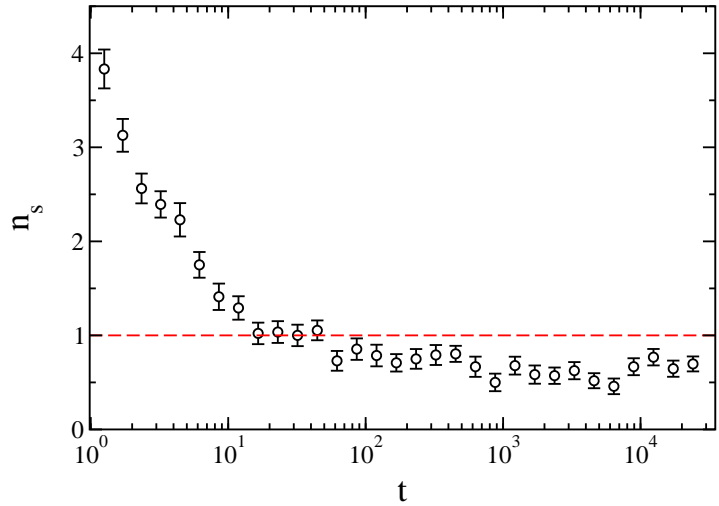}
\caption{Time behavior of the mean order $n_s$ of the {\it saddles}.}
\label{fig2}
\end{figure}

\begin{figure}[hbt]
\centering
\includegraphics[width=.5\textwidth,angle=-90]{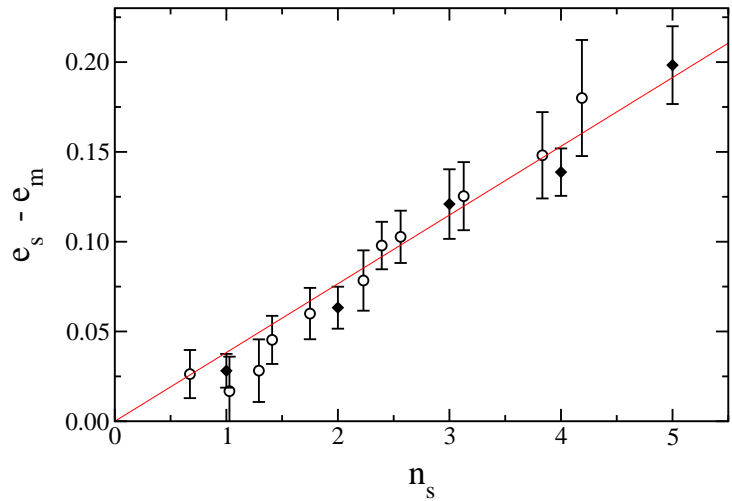}
\caption{The energy elevation of {\it saddles} from the underlying minima as a function of saddle order $n_s$.
The open circles are the data obtained from Fig.1 and Fig.2 (the minima are defined 
starting from instantaneous configurations), while the open diamonds are obtained  determining
the underlying minima starting from saddle configurations.
The fit line determines the energy barrier (expressed as total energy) 
$\Delta E = 10$.}
\label{fig3}
\end{figure}

\vspace{-1.cm}

\begin{figure}[hbt]
\centering
\includegraphics[width=.5\textwidth,angle=-90]{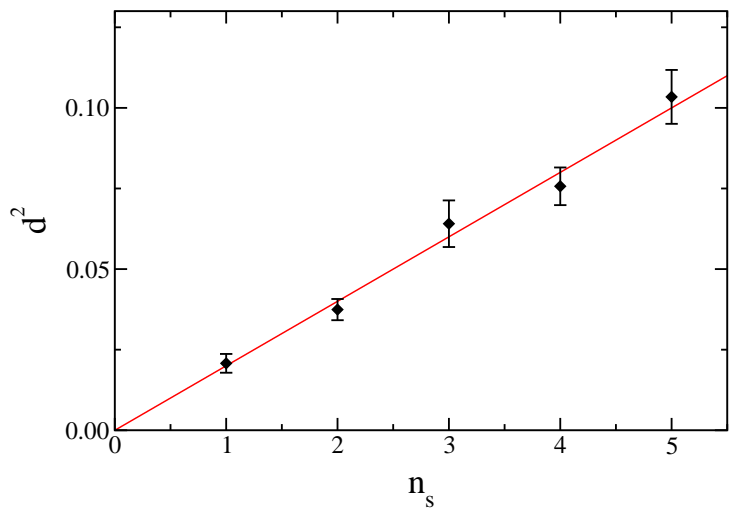}
\caption{Mean square distance $d^2$ between adjacent minima vs. {\it saddles} order $n_s$.
The line is a linear fit, with slope $d^2_0 = 0.02$.}
\label{fig4}
\end{figure}

\end{document}